\documentclass[twocolumn,showpacs,preprintnumbers,amsmath,amssymb]{revtex4}
\usepackage{graphicx,amsmath,amssymb}

\begin{document}

\title{The dependence of transverse and longitudinal resolutions on incident Gaussian beam widths in the illumination part of optical scanning microscopy}

\author{Hyung-Su Chon, Gisung Park, Sang-Bum Lee, Seokchan Yoon, Jaisoon Kim, Jai-Hyung Lee, and Kyungwon An}
\email{kwan@phya.snu.ac.kr}

\address{School of Physics, Seoul National University, Seoul 151-747, Korea}

\begin{abstract}
We studied both theoretically and experimentally the intensity distribution of a Gaussian laser beam when it was focussed by an objective lens with its numerical-aperture (NA) up to 0.95. Approximate formulae for full widths at half maximum (FWHM) of the intensity distribution at focus were derived for very large and very small initial beam waists with respect to the entrance pupil radius of the objective lens. In experiments the energy flux through a 0.5 micron pinhole was measured for various pinhole positions. We found that the FWHM's at focus in the transverse and the longitudinal directions do not increase much from the ultimate FWHM's until the input beam waist is reduced below the half of the entrance pupil radius. In addition, we observed significance of the spatial distribution of the input beam against a true Gaussian beam profile in the case of small initial beam waist. For high NA with resulting focal beam waists comparable to or smaller than the wavelength of the laser, the interaction between the electric field and the conducting surface of the pinhole caused the transverse FWHM to be measured slightly smaller than FWHM of the unperturbed intensity distribution convoluted with the pinhole opening.
\end{abstract}

\maketitle

\section{Introduction}

The spatial resolution in optical scanning microscopy is critically dependent on both the beam spot size near the focus of a scanning objective lens and how the focal spot is imaged back onto an imaging plane. In order to achieve the ultimate resolution the beam spot size at focus should be minimized for a given illumination source. Otherwise, the spatial resolution is degraded and it cannot be recovered however well one handles the imaging of the focal spot. For proper accessing the minimal focal beam spot, one should be able to calculate and measure the beam spot size accurately. 

In many experiments using an objective lens we usually assume that the incident beam is a plane wave apertured by the entrance pupil of the objective lens. However, the light source in the optical scanning microscopy is often a Gaussian laser beam, not an ideal plane wave. One can expand the Gaussian beam and let the central part of it, simulating a plane wave, incident on the objective lens. A practical question is then how large the beam should be expanded with respect to the entrance pupil size of the objective lens in order to obtain a spatial resolution comparable to that with the ideal plane wave input. 

To answer this question, we need to know the near-focal plane intensity distribution of a Gaussian laser beam with an initial beam width $w_0$ when focused by an objective lens with an entrance pupil diameter $D$ \cite{Marshall91,Belland82,Tanaka85}. The intensity distribution, in general, can be calculated by the electromagnetic diffraction theory of Richards and Wolf \cite{Wolf59,Richards59,Mansuripur89}. This theory is based on the vectorial equivalent of the Kirchhoff-Fresnel integral in the Debye approximation \cite{Born99,Stamnes86}. 

The intensity distribution in the region of focus have been measured in several experiments by using a knife-edge \cite{Schneider81,Firester77,Quabis01,Dorn03} and a tapered fiber \cite{Rhodes89,Rhodes02}. However, a systematic investigation of the near-focus intensity distribution in the {\em non-paraxial} regime as a function of the input Gaussian beam width $w_0$ has not been reported. 

In the present work, we re-examine the diffraction theory of Richards and Wolf for input Gaussian beams.  In two limiting cases of very small and very large input beam widths, we derive approximate formula for the full width at half maximum (FWHM) of the intensity distribution at focus in the longitudinal and transverse directions. We then confirm the validity of theoretical predictions in actual experiments employing objective lenses with numerical apertures of 0.4, 0.75 and 0.95 for various Gaussian input beam widths.

This paper is organized as follows.  In Sec.\ \ref{sec2}, we first theoretically examine transverse and longitudinal FWHM's near the focal plane for an arbitrary input beam waist $w_0$ and then derive approximate formulae for limiting cases, $w_0\ll R$ and $w_0\gg R$. Experiment is described in Sec.\ \ref{sec3} and results and discussion are presented in Secs.\ \ref{sec4} and \ref{sec5}. We summarize the work in Sec.\ \ref{sec6}.

\section{Theory} \label{sec2} 

Suppose a Gaussian beam with a waist $w_0$ is incident on an objective lens with a high NA and an entrance pupil radius of $R$. We can think of three different regimes, namely, (i) $w_0 \ll R$, (ii) $w_0 \sim R$,  and (iii)$w_0 \gg R $. We first consider a general theory which can address all three regimes and then discuss regimes (i) and (iii) as limiting cases of the general theory.

\subsection{Field distribution near the focal region in general cases}

We use the electromagnetic diffraction theory of Richards and Wolf \cite{Wolf59,Richards59,Mansuripur89} for the numerical calculation of the intensity distribution of the focused beam. For the integral, we choose our Cartesian coordinate system in the following way (see Fig.\ 1). The origin is located at the focus, $z$ axis coincides with the optic axis of the optical system under consideration, pointing in the beam propagation direction and $x$ axis points in the polarization direction of the incident field $\bf{e}_0$. A time-independent part $\bf{e}(\bf{r})$ of the analytical solution of the Helmholtz's equation for the electric field at a point P$(\bf{r})$ in the image space of our optical system is given by $^{1}$

\begin{figure}
\includegraphics[width=3.4in]{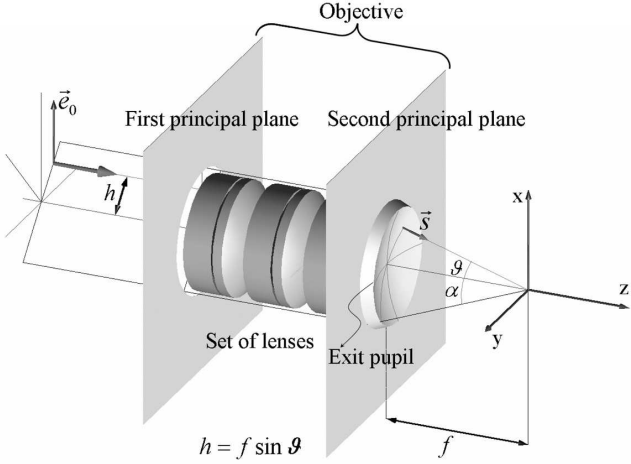}
\caption{Coordinate system for the calculation of the intensity distribution in the region of focus.}
\label{fig1}
\end{figure}

\begin{equation}
\mathbf{e} (\mathbf{r})=-\frac{ikf}{2\pi} \int\!\!\!\int_{\Omega}
\frac{\mathbf{a}(s_{x},s_{y})}{s_{z}} \exp\{ik[\Phi(s_{x},s_{y})+ \mathbf{s} \cdot \mathbf{r} ]\}\,ds_{x}\,ds_{y},
\label{eq1}
\end{equation}
where $\mathbf{s}=(s_x,s_y,s_z)$  is a unit vector pointing in the direction of a ray, $ \Phi(s_x,s_y)$ represents aberration in the optical system, $\Omega$ is the solid angle subtended by the exit pupil of the objective lens from the origin, the focus, and $\bf{a}$, called an electric strength factor, is the electric field incident on the exit pupil after passing through the lens. Similarly, the magnetic field $\bf{h}(\bf{r})$ can be written in the same way in terms of a different strength factor $\mathbf{b}(=\mathbf{s} \times \mathbf{a})$. Eq.(\ref{eq1}) is valid only if $ kf \gg 1$, where $f$ is the focal length.

We introduce spherical polar coordinates $(f,\vartheta,\varphi)$ for the point Q on the exit pupil and $(r,\theta,\phi)$ for the observation point P in the image space. The Cartesian components of the strength vector $\bf{a}$ can then be written as

\begin{eqnarray}
a_x &=& e_0 (\vartheta) \sqrt{\cos \vartheta}
[\cos \vartheta + \sin^2 \varphi(1-\cos \vartheta)],\nonumber\\
a_y &=& e_0 (\vartheta) \sqrt{\cos \vartheta}(\cos \vartheta
-1)\cos \varphi \sin \varphi, \nonumber\\
a_z &=& -e_0 (\vartheta) \sqrt{\cos \vartheta} \sin \vartheta \cos
\varphi\;,
\label{eq2}
\end{eqnarray}
where $e_0(\vartheta)$ is the amplitude of the incident electric field $\bf{e}_0$. Similar expressions hold for the components of the magnetic field strength factor $\bf{b}$. On substitution of Eq.(\ref{eq2}) into Eq.(\ref{eq1}) with $\bf{s}=(\sin\vartheta \cos\varphi,\sin\vartheta \sin\varphi,\cos\vartheta)$, we obtain the following expressions for the Cartesian components of $\bf{e}$.

\begin{eqnarray}
e_x (\mathbf{r})&=&-\frac{i}{2} kf(I_0 + I_2 \cos 2
\phi), \nonumber\\
e_y (\mathbf{r})&=&-\frac{i}{2} kf I_2 \sin 2 \phi, \nonumber \\
e_z (\mathbf{r})&=&-i kf I_1 \cos \phi,
\label{eq3}
\end{eqnarray}
where

\begin{eqnarray}
I_0(r,\theta)&=&\int_{0}^\alpha e_0(\vartheta)\sqrt{\cos
\vartheta} \sin \vartheta (1 + \cos \vartheta) \nonumber\\
 & & \;\;\;\;\;\;\times J_0(kr\sin
\vartheta \sin \theta) \exp(ikr\cos \vartheta \cos
\theta)\,d \vartheta,\nonumber\\
I_1(r,\theta)&=&\int_{0}^\alpha e_0(\vartheta)\sqrt{\cos
\vartheta} \sin^2 \vartheta \nonumber\\
& & \;\;\;\;\;\;\times J_1(kr\sin \vartheta \sin
\theta) \exp(ikr\cos \vartheta \cos \theta) \,d \vartheta ,\nonumber \\
I_2(r,\theta)&=& \int_{0}^\alpha e_0(\vartheta)\sqrt{\cos
\vartheta} \sin \vartheta (1-\cos \vartheta) \nonumber\\
& &\;\;\;\;\;\;\times J_2(kr\sin \vartheta \sin \theta)
\exp(ikr\cos \vartheta \cos \theta)\,d \vartheta, \label{eq4}
\end{eqnarray}
where $\alpha$ is a semi-aperture angle satisfying $\Omega=2\pi(1-\cos\alpha)$ and its Sine value is the numerical aperture (NA=$\sin \alpha $).

For a well-collimated Gaussian beam with a beam waist $w_0$ and an amplitude $A_0$, $e_0(\vartheta)$ can be written as

\begin{equation}
e_0(\vartheta)= A_0 \exp[-(f \sin \vartheta /w_0)^2].
\label{eq5}
\end{equation}
under the Abbe's sine condition \cite{mansuripur02}. 

The quantity to be measured in our experiment to be presented below is the power transmitted by a small aperture near the focal plane. This quantity is nothing but the time-averaged $z$-component of the Poynting vector, which is given by
\begin{equation}
S_z (\mathbf{r})= {c (kf)^2 \over {32\pi}} (|I_0|^2 - |I_2|^2),
\label{eq6}
\end{equation}
where $c$ denotes the speed of light in vacuum.

\subsection{Large beam waist limit, $w_0 \gg R $}
Since $w_0 \gg R$, we can approximate the incident Gaussian beam as a plane wave and use the results in the previous section with a substitution $ e_0(\vartheta)= A_0 (constant)$ in  Eq.\ (\ref{eq4}).

\subsubsection{Transverse spot size ($ \Delta x_{\rm FWHM} $)}

\begin{figure}
\includegraphics[width=2.5in]{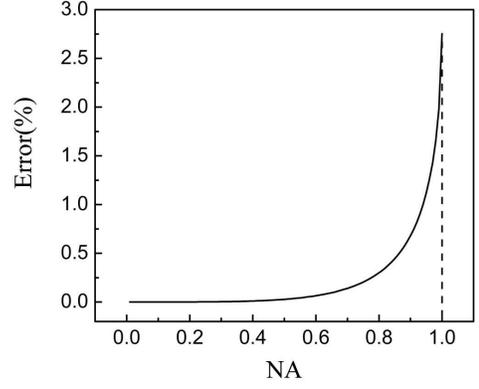}
\caption{Error in $\Delta x_{\rm FWHM}$ approximated by Eq.\ (\ref{eq9}) with respect to the exact one by Eqs.\ (\ref{eq4}) and (\ref{eq6}) as a function of NA.}
\label{fig2}
\end{figure}

The field distribution in the focal plane of the objective lens can be written as

\begin{eqnarray}
I_0 (r,\theta = \pi/2)&=& A_0 \int_{0}^\alpha \sqrt{\cos
\vartheta}
\sin \vartheta (1 + \cos \vartheta) \nonumber\\
& & \;\;\;\;\;\;\times J_0(kr\sin \vartheta)\,d \vartheta,\nonumber\\
I_1 (r,\theta = \pi/2)&=& A_0 \int_{0}^\alpha \sqrt{\cos
\vartheta} \sin^2 \vartheta \nonumber\\
& & \;\;\;\;\;\;\times J_1(kr\sin \vartheta ) \,d \vartheta , \nonumber\\
I_2  (r,\theta = \pi/2)&=& A_0 \int_{0}^\alpha \sqrt{\cos
\vartheta}
\sin \vartheta (1-\cos \vartheta)\nonumber\\
& & \;\;\;\;\;\;\times J_2(kr\sin \vartheta )\,d \vartheta,
\label{eq7}
\end{eqnarray}
In general, $I_0 \gg I_1, I_2$ and thus the transverse spot size at focus is mostly determined by $I_0$ integral. Further approximation is then obtained by noting that the functional factor $(1+\cos \vartheta)/2$ is approximately equal to $\sqrt{\cos\vartheta}$, which can be easily verified by Taylor series expansion of these two. This approximation is reasonably good even when $\vartheta \simeq 1$. For example, the difference between these two fuctional factors is 4.8\% for $\vartheta=1$. Under this approximation, Eq.\ (\ref{eq7}) becomes

\begin{eqnarray}
I_0 &\approx& 2A_0 \int_{0}^\alpha \cos \vartheta \sin \vartheta J_0(kr\sin \vartheta)\,d \vartheta \propto  \frac{J_1(kr \sin\alpha)}{kr\sin\alpha}\;,  \nonumber\\
\label{eq8}
\end{eqnarray}
which is of the same form as the Fraunhofer diffraction by a circular aperture. Although the paraxial assumption $\sin \alpha \ll 1$ is used in the Fraunhofer diffraction theory, our approximate result, Eq.\ (\ref{eq8}), is still applicable to non-paraxial cases with $\alpha$ up to the order of unity. This finding is new and has not been recognized. The transverse spot size is then obtained from Eq.\ (\ref{eq8}) as
\begin{equation}
\Delta x_{\rm FWHM}\simeq \frac{2\times 1.6163}{k \sin \alpha}=
0.5145\frac{\lambda}{\rm NA}
\label{eq9}
\end{equation}
Figure \ref{fig2} shows the difference between $\Delta x_{\rm FWHM}$ approximated by Eq.\ (\ref{eq9}) and the exact one by Eqs.\ (\ref{eq4}) and (\ref{eq6}). The approximation is excellent in that the difference is as small as 2.8\% even when NA=1, the largest possible NA value.

\subsubsection{Longitudinal spot size ($ \Delta z_{\rm FWHM} $)}

\begin{figure}
\includegraphics[width=2.5in]{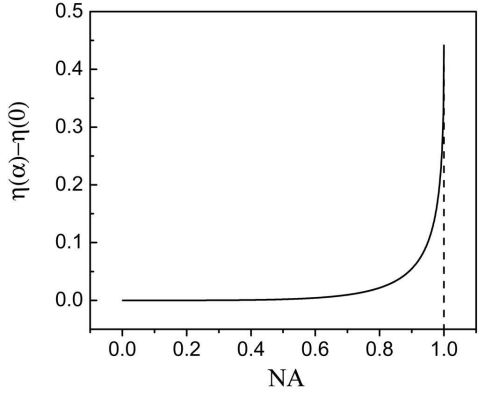}
\caption{Numerical factor $\eta(\alpha)$ in Eq.\ (\ref{eq11}).} 
\label{fig3}
\end{figure}

The field distribution in the $z$-axis near the focus is given by
\begin{eqnarray}
I_0(r=z,\theta = 0)&=& A_0 \int_{0}^\alpha \sqrt{\cos \vartheta}
\sin
\vartheta (1 + \cos \vartheta) \nonumber\\
& &\;\;\;\;\;\; \times \exp(ikz\cos \vartheta )\,d \vartheta,\nonumber\\
I_1(r,\theta = 0)&=& I_2(r,\theta = 0)= 0,
\end{eqnarray}
Under the same approximation as above,
\begin{eqnarray}
I_0&\approx& 2A_0 \int_{0}^\alpha \cos \vartheta \sin\vartheta \exp(ikz\cos \vartheta ) d \vartheta \nonumber\\
&=&\frac{2A_0}{(kz)^2}\int_{kz \cos\alpha}^{kz} q \exp(iq) dq \nonumber\\
&\propto& \left(\sin^2 \alpha \right)\left[ \left(\frac{\sin
x}{x}\right)-i \tan^2 \frac{\alpha}{2}\left(\frac{x \cos x - \sin
x}{x^2}\right) \right] \nonumber\\
\label{eq10}
\end{eqnarray}
where $x=kz \sin^2 (\alpha/2)$. For $\alpha$ up to unity, the contribution from the second term in $|I_0|^2$ is negligibly small, proportional to $\tan^4\frac{\alpha}{2} < 0.089$, and thus $|I_0|^2$ is approximately given by the Sinc function squared, which is again the same as the Fraunhofer diffraction result except that $x$ is proportional to $\alpha^2$ not to $\sin^2(\alpha/2)$ in the usual Fraunhofer diffraction. For an arbitrary $\alpha$, $\Delta z_{\rm FWHM}$ is obtained from Eq.\ (\ref{eq10}) as
\begin{equation}
\Delta z_{\rm
FWHM}=\frac{\eta(\alpha)\lambda}{4\sin^2\frac{\alpha}{2}}=\frac{\eta(\arcsin
{\rm NA})\lambda}{4\sin^2\ (\frac{1}{2}\arcsin {\rm NA})}
\label{eq11}
\end{equation}
where the slowly varying function $\eta(\alpha)$ is plotted in Fig.\ \ref{fig3}. For $\alpha$ up to unity, we can approximate $\eta(\alpha)\simeq \eta(0)\simeq 1.772$, by which our error is only 1.7\% for $\alpha=1$ and 5.4\% for $\alpha=1.25$, which corresponds to NA=0.95. Under this approximation,

\begin{equation}
\Delta z_{\rm FWHM}\simeq
\frac{1.772\lambda}{4\sin^2\frac{\alpha}{2}}=\frac{1.772\lambda}{4\sin^2\
(\frac{1}{2}\arcsin {\rm NA})}
\label{eq12}
\end{equation}
which reduces to the usual Fraunhofer diffraction result
\begin{equation}
\Delta z_{\rm FWHM}\simeq 1.772\frac{\lambda}{\alpha^2}\simeq
1.772\frac{\lambda}{{\rm NA}^2}\;,
\label{eq13}
\end{equation}
under the paraxial condition, $\alpha\ll 1$.

\subsection{Small beam waist limit, $w_0 \ll R $}

Although the numerical aperture of the lens is assumed to be large, only the central portion of the objective lens is utilized by the incident Gaussian beam when $w_0 \ll R$. One can define an effective numerical aperture $\rm NA_{eff}$ as ${\rm NA_{eff}}\equiv w_0/f \ll 1$, and thus the paraxial approximation can be effectively applied. One is allowed to use Gaussian optics to calculate the beam size in the focal region. Particulary, when the incident beam has a minimum waist at the entrance pupil of the objective lens, the Gaussian optics provides a simple formula for the field distribution in the region of focus.

\subsubsection{Transverse spot size ($ \Delta x_{\rm FWHM} $)}

 The Gaussian beam waist ${w_0}'$ in the region of focus is given by
\begin{eqnarray}
{w_0}'={{f \lambda} \over {\pi w_0}}
\label{eq14}
\end{eqnarray}
where $w_0$ is the minimum beam waist of the incident beam located at the entrance pupil of the objective lens. The above $1/e$-width can be converted to a full width at half maximum as
\begin{equation}
\Delta x_{\rm FWHM} = 2 \sqrt{\ln{\sqrt{2}}}\;w_0' \simeq  0.375
\frac{\lambda}{\rm NA_{eff}}\;,
\label{eq15}
\end{equation}
where ${\rm NA_{eff}}\equiv w_0/f$.

We can also derive the above result from the $I$ integrals for general cases. From Eq.\ (\ref{eq4}), the field distribution in the focal plane can be written as
\begin{eqnarray}
I_0(r,\theta=\pi/2)&=&\int_{0}^\alpha e_0(\vartheta)\sqrt{\cos
\vartheta} \sin \vartheta (1 + \cos \vartheta) \nonumber\\
 & & \;\;\;\;\;\;\times J_0(kr\sin
\vartheta) \,d \vartheta,\nonumber\\
I_1(r,\theta=\pi/2)&=&\int_{0}^\alpha e_0(\vartheta)\sqrt{\cos
\vartheta} \sin^2 \vartheta \nonumber\\
& & \;\;\;\;\;\;\times J_1(kr\sin \vartheta) \,d \vartheta , \nonumber\\
I_2(r,\theta=\pi/2)&=& \int_{0}^\alpha e_0(\vartheta)\sqrt{\cos
\vartheta} \sin \vartheta (1-\cos \vartheta) \nonumber\\
& &\;\;\;\;\;\;\times J_2(kr\sin \vartheta) \,d \vartheta,
\label{eq16}
\end{eqnarray}
where $e_0(\vartheta)$ is given by Eq.\ (\ref{eq5}). Since $e_0(\vartheta)$ is significant only when $\sin\vartheta \le w_0/f \ll 1$, the integrands above count only when $\vartheta \ll 1$, and thus we can rewrite the above as
\begin{eqnarray}
I_0&\approx&2\int_{0}^\alpha e_0(\vartheta)\vartheta J_0(kr\vartheta) \,d \vartheta,\nonumber\\
I_1&\approx&\int_{0}^\alpha e_0(\vartheta)\vartheta^2 J_1(kr \vartheta) \,d \vartheta , \nonumber\\
I_2&\approx& \frac{1}{2}\int_{0}^\alpha e_0(\vartheta) \vartheta^3
J_2(kr \vartheta) \,d \vartheta,
\label{eq17}
\end{eqnarray}
Since $I_1/I_0 \sim (w_0/f)^2 \ll 1$ and $I_2/I_0 \sim (w_0/f)^4 \ll 1$, the field distribution is mostly determined by $I_0$. We can further simply the $I_0$ integral as
\begin{eqnarray}
I_0 &\propto &\int_{0}^\alpha \exp[-\left(f\vartheta /w_0 \right)^2]\vartheta J_0(kr\vartheta) \,d \vartheta\nonumber\\
&\propto& \int_{0}^{f\alpha/w_0} \exp(-x^2)x J_0 \left(\frac{k r w_0}{f} x\right) \,d x \nonumber\\
&\simeq& \int_{0}^{\infty} \exp(-x^2)x J_0 \left(\rho x\right) \,d
x = \exp[-(\rho/2)^2] \label{eq18}
\end{eqnarray}
where $\rho=krw_0/f$, from which we obtain an $1/e$ width of the field distribution as $2f/kw_0$, which is nothing but $w_0'$ in Eq.\ (\ref{eq14}).\\

\subsubsection{Longitudinal spot size ($ \Delta z_{\rm FWHM} $)}

In Gaussian optics, the Rayleigh range ${z_0}'$ in the region of focus is given by
\begin{eqnarray}
{z_0}' = {{\pi {{w_0}'}^2}\over {\lambda}} ={\lambda \over \pi}
\left({f \over {w_0}} \right)^2 .
\label{eq19}
\end{eqnarray}
The FWHM value in the $z$ direction is just twice of the Rayleigh range. 
\begin{eqnarray} \Delta z_{\rm FWHM}= 2
\left(\frac{\lambda}{\pi}\right) \left(\frac{f}{w_0} \right)^2
\simeq 0.6366\frac{\lambda}{\rm NA_{eff}^2}. \label{eq20}
\end{eqnarray}
We can also derive Eq.\ (\ref{eq20}) from Eq.\ (\ref{eq4}):
\begin{eqnarray}
I_0(r,\theta=0)&=&\int_{0}^\alpha e_0(\vartheta)\sqrt{\cos \vartheta} \sin \vartheta (1 + \cos \vartheta) \nonumber\\
 & & \;\;\;\;\;\;\times \exp(ikr \cos\vartheta)\,d \vartheta,\nonumber\\
I_1(r,\theta=0)&=&0=I_2(r,\theta=0). \label{eq21}
\end{eqnarray}
Again, the integrand is significant only when $\vartheta\le w_0/f \ll 1$, and thus
\begin{eqnarray}
I_0 &\propto&\int_{0}^\infty \exp(-x^2) x \exp\left\{ikr\left[1-\frac{1}{2}\left(w_0 x/f\right)^2\right]\right\}dx \nonumber\\
&=&\frac{1}{2}\exp(ikr)\int_{0}^\infty \exp(-q) \exp\left[-\frac{i}{2}kr (w_0/f)^2 q\right]dq \nonumber\\
&\propto& \left(1+i\frac{krw_0^2}{2f^2}\right)^{-1}\;,
\label{eq22}
\end{eqnarray}
and thus the intensity distribution is proportional to a Lorentzian
\begin{equation}
\left|I_0 \right|^2
\propto\frac{1}{r^2+\left(\frac{2f^2}{kw_0^2}\right)^2}\;,
\label{eq23}
\end{equation}
from which we obtain $\Delta z_{\rm FWHM}=4f^2/kw_0^2$ identical to the one in Eq.\ (\ref{eq20}).

\subsection{Application to NA=0.4, 0.75 and 0.95}
\begin{figure}
\includegraphics[width=3.4in]{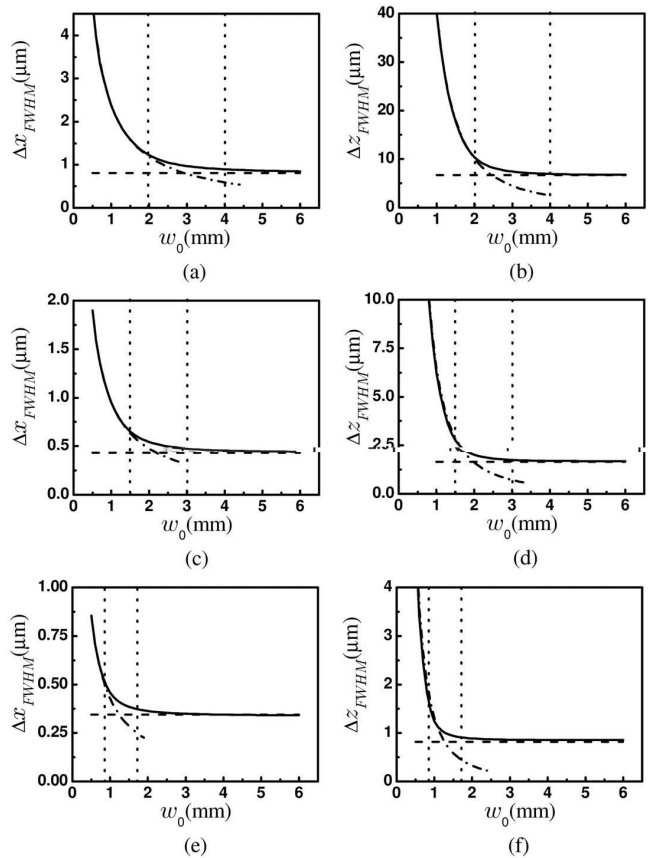}
\caption{Dependence of transverse and longitudinal FWHM values, $\Delta x_{\rm FWHM}$ and $\Delta z_{\rm FWHM}$, respectively, on $w_0$ of the incident Gaussian beam. Vertical dotted lines indicate $w_0=R/2$ and $R$. (a)-(b) NA=0.4, (c)-(d) NA=0.75, and (e)-(f) NA=0.95.}
\label{fig4}
\end{figure}

In Fig.\ \ref{fig4}, theoretical $\Delta x_{\rm FWHM}$ and $\Delta z_{\rm FWHM}$ values for NA=0.4, 0.75, 0.95, respectively, are plotted. The solid lines represent FWHM's calculated from Eqs.\ (\ref{eq4}) and (\ref{eq6}).  The dash-dotted lines in Figs.\ \ref{fig4}(a),(c) and (e) are calculations based on Eqs.(\ref{eq15}) and those in Figs.\ \ref{fig4}(b),(d) and (f) are given by Eq.\ (\ref{eq20}). Similarly, the dashed lines are calculated from Eqs.\ (\ref{eq9}) and (\ref{eq12}). These dash-dotted and dashed lines represent two limiting cases, $w_0/R\ll 1$ and $w_0/R \gg 1$, respectively, of the general curves which are valid for any $w_0/R$ values for given NA's.

\section{Experiment} \label{sec3}

\subsection{Quantity to be measured}

In order to measure the energy flux or the $z$-component of the Poynting vector associated with the field distribution near the focal plane, we place a sub-micron pinhole at various positions and measure the light power transmitted by the pinhole. By scanning the pinhole transversely and longitudinally, we can map out the distribution of the energy flux. Since the pinhole is made of a conductor, the field distribution near the pinhole is slightly modified. However, we assume that the effect of the interaction between the pinhole and the field on the measurement of the energy flux is negligible. The validity of this assumption will be discussed in the next section.  

The resolution of an optical microscope is determined by the electric field distribution in the focal region of the objective lens since samples response to the electric field of an illumination light \cite{E-field-resp}.  In our experiment, however, we measure the energy flux or the $z$-component of the Poynting vector associated with the electric field distribution as mentioned above. According to our theoretical investigation, the smallest FWHM of the electric field distribution is about the same as that of the $z$-component distribution of the Poynting vector $S_z$ in the focal region up to NA $\simeq 1$ under our experimental conditions. The difference between those two FWHM's is about 8\%, except for the usual difference,  {\em i.e.}, the detailed structure in the electric field distribution elongated in the incident polarization direction \cite{Richards59}. Therefore, the transverse and longitudinal FWHM's of $S_z$ distribution well approximate those of the electric field distribution and thus they can be used as measures of the optical resolution associated with the illumination part of an optical scanning microscope.

\subsection{Experimental Setup} 

\begin{figure}
\includegraphics[width=3.4in]{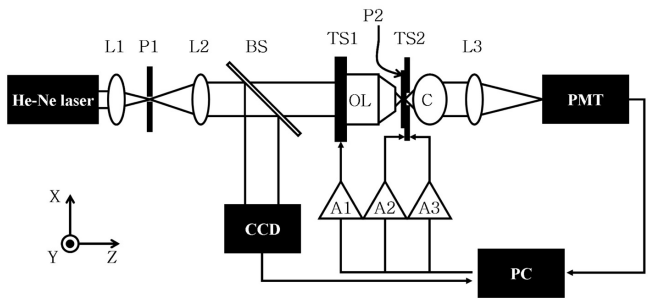}
\caption{Experimental setup for measuring the profile of the beam focused by an objective lens. L1, L2, L3: lenses, BS: beam splitter, TS1: translation stage driven by a closed-loop-feedback stepper motor, TS2: translation stage driven by closed-loop-feedback PZT actuators, OL: objective lens, CCD: charge-coupled device detector, P1, P2: pinholes, C: condenser, PMT: photomultiplier tube, and A1, A2, A3: scan control voltage signals from an analog-digital converter board on a personal computer. Signal A1 controls the $z$ translation of the objective lens and  signals A2 and A3 control the $x$, and $y$ translation of the pinhole stage. A spatial filter is formed by L1, P1, and L2.}
\label{fig5}
\end{figure}

Our experimental setup is shown in Fig.\ \ref{fig5}. A He-Ne laser (632.8nm) with $x$-polarization was first incident on a spatial filter, and then expanded and collimated to a Gaussian beam with a beam waist $w_0$. Its profile was measured by a motorized beam profiler. An objective lens was mounted on a $xyz$-translation stage with its $z$ coordinate scanned by a step motor in a closed feedback loop (model M-126.PD from PI Ltd.) and thus it could be coarse-positioned manually and fine-scanned by the step motor with 0.125 $\mu$m resolution in the $z$ direction.

Infinity-corrected microscope objective lenses with NA=0.4, 0.75 (both from NIKON) and 0.95 (from OLYMPUS), respectively, were used. The spherical aberration coefficients of the objective lenses were measured with a Twymann-Green interferometer (Zygo) and the results are 0.44$\lambda$, 0.21$\lambda$, 0.43$\lambda$ for NA=0.4, 0.75, 0.95, respectively. In FWHM measurement to be presented below the error caused by these values of spherical aberration is estimated to be negligible, as small as 0.1\% or less.

\begin{figure}
\centering
\includegraphics[width=2.5in]{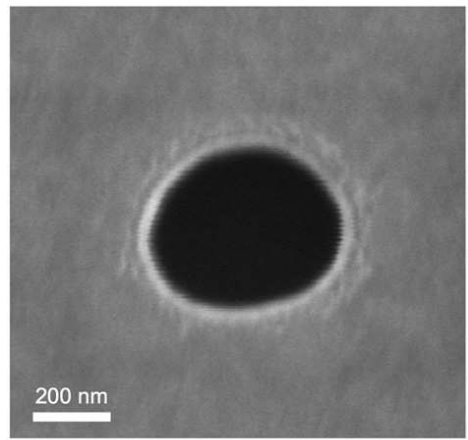}
\caption{Scanning electron microscopy image of the pinhole ($\phi=0.5 \pm 0.05 \mu$m) used as an intensity probe in our experiment.} 
\label{fig6}
\end{figure}

A pinhole (see Fig.\ \ref{fig6}) with a diameter of (0.50 $\pm$ 0.05) $\mu$m, which served as an intensity probe, was mounted on a translation stage driven by piezoelectric transducer (PZT) stacks in a closed feedback loop for scanning in the $x$- and $y$ directions. Typical stroke errors of these PZT stacks were less than 0.1\% of their stroke ranges.

The pinhole was made with the electron-beam etching technique on a thin Ti:sapphire substrate with a gold coating layer of 200 nm thickness. The gold layer in a circle of 0.5 micron diameter was removed to form a pinhole.

The light transmitted through the pinhole was detected by a photomultiplier tube and the signal was digitized by a data acquisition board in a computer as a function of the pinhole position. A resulting image amounted to a 200$\times$200 array of pixels.

\section{Results} \label{sec4}

We measured the intensity distribution for a Gaussian beam with an initial beam waist of $w_0$=0.57, 0.97, 1.59, 2.57, 3.1, 3.58, and 5.88 mm. From the measured intensity distribution in the $xz$ meridional plane, we determined FWHM's in the $x$ direction ($\Delta x_{\rm FWHM}$) and in the $z$ direction($\Delta z_{\rm FWHM}$).

\begin{figure}
\centering
\includegraphics[width=3.4in]{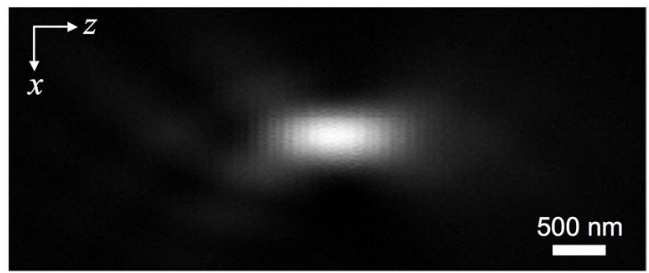}
\caption{Observed $xz$-profile in the focal region for NA=0.95 objective lens. The image covers a scan area of 2.5 $\mu$m $\times$ 6 $\mu$m.} 
\label{fig7}
\end{figure}

For instance, the intensity profile created by an objective lens with NA=0.95 for an input beam of $w_0$=5.88 mm is shown in Fig.\ \ref{fig7}. Since the entrance pupil radius $R$ of the objective lens was 1.71 mm, we can consider the incident beam as a plane wave. The $x$-$z$ profile corresponded to an actual area of 2.5 $\mu$m $\times$ 6 $\mu$m. The measured $x$- and $z$-FWHM values were 0.4 $\mu$m and 1.03 $\mu$m, respectively.

The time-averaged $z$ component of the Poynting vector in the near focus was calculated from Eq.\ (\ref{eq6}). To compare experiment with theory, we assumed that the total amount of light detected by the PMT through the pinhole was proportional to the convolution of the $z$ component of the Poynting vector with the pinhole opening.
\begin{equation}
\tilde {S}_z (x,y) = \int\int S_z (x',y') P(x-x',y-y')\,d x' \, d y'
\label{eq24}
\end{equation}
where $P(x,y)$ is an aperture function for the pinhole. This assumption is equivalent to saying that the possible field distortion by the conducting surface of the pinhole substrate does not affect the amount of energy flow through the pinhole much so that we just integrate the surface-normal component of the unperturbed Poynting vector calculated for the absence of the pinhole over the aperture function of the pinhole.

\begin{figure}
\includegraphics[width=3.4in]{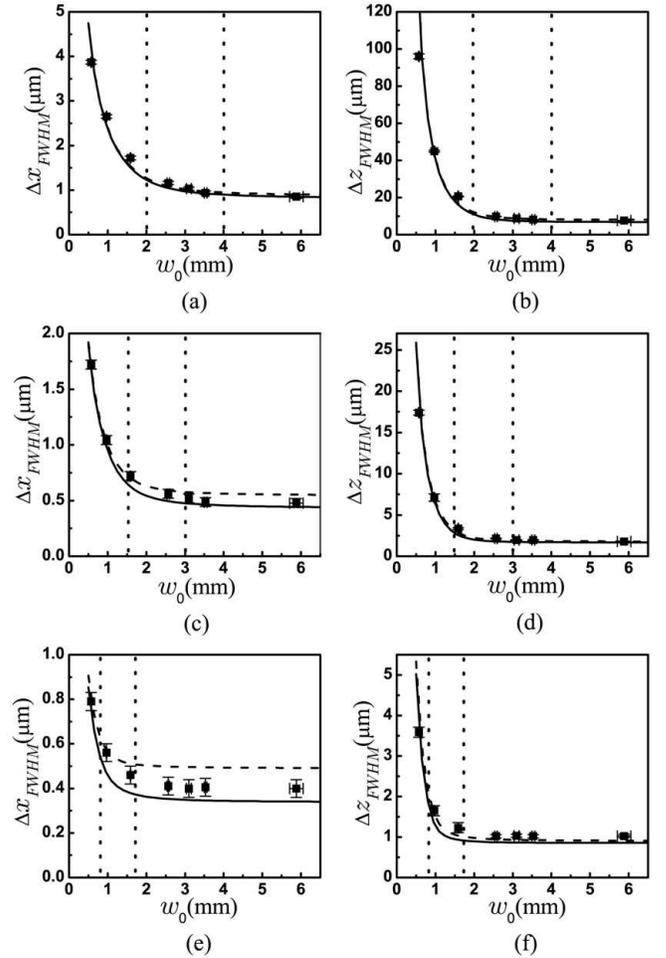}
\caption{Dependence of transverse ($x$) and longitudinal ($z$) FWHM values on $w_0$ of the incident Gaussian beam. Unconvoluted FWHM's obtained from Eq.\ (\ref{eq6}) are represented by solid lines whereas the convoluted FWHM's given by Eq.\ (\ref{eq24}) are drawn as dashed lines. Experimental results are marked by square dots with error bars. 
Independently measured spherical aberrations were included in the calculations. Vertical dotted lines indicate $w_0=R/2$ and $R$. (a)-(b): NA=0.4, (c)-(d): NA=0.75, and (e)-(f): NA=0.95.}
\label{fig8}
\end{figure}

The dependence of the measured $x$- and $z$-FWHM values on the input Gaussian waist $w_0$ for NA=0.4, 0.75 and 0.95 are shown in the Fig.\ \ref{fig8}, where (a) and (b) are for NA=0.4, (c) and (d) for NA=0.75 and (e) and (f) for NA=0.95, respectively. FWHM values obtained from Poynting vector $S_z(x,y)$ and convoluted Poynting vector $\tilde{S}_z(x,y)$ are represented by solid and dashed lines, respectively, and experimental results are represented by square dots. The spherical aberration of the object lenses was included in the calculation of $S_z$. The agreement between experiment and theory is reasonably good. 

The smallest ($x$-FWHM,  $z$-FWHM) values measured in the experiment are (0.86 $\mu$m, 7.62 $\mu$m) for NA=0.4, (0.48 $\mu$m, 1.79 $\mu$m) for NA=0.75 and (0.40 $\mu$m, 1.03 $\mu$m) for NA=0.95. These values are in good agreement with the convoluted FWHM values except for the $x$-FWHM values for NA=0.75 and 0.95 (see Figs.\ \ref{fig8}(c) and (e)), for which the observed FWHM is slightly smaller than the convoluted FWHM but larger than the un-convoluted FWHM.

\section{Discussions} \label{sec5}
\subsection{Effect of the interaction between the pinhole and the electric field}

The experimental results summarized in Fig.\ \ref{fig8} show that the observed FWHM is smaller than the FWHM of the pinhole-convoluted $\bar S_z$ distribution when the focused beam spot size is comparable to or smaller than the pinhole size. For such small focal beam spots, the pinhole seems to behave as a smaller pinhole for light transmission. This phenomenon appears to be caused by the interaction of the electric field and the conducting surface of the pinhole. The distortion of the field distribution near a conducting structure like a pinhole is usually in the sub-wavelength scale and thus it can be neglected if the range of the field distribution is much larger than the wavelength. If the range of the field distribution is in the sub-wavelength scale, as in the case of $x$-FWHM for NA=0.75 and 0.95, the field distortion effect could be non-negligible.
Our numerical simulation supports this reasoning. Nonetheless, the effect is still small, amounting to at most 20\% with respect to the simple-minded convoluted FWHM, under our experimental conditions and thus our previous assumption of neglecting this effect could be well justified in the first-order approximation.

\subsection{Range of input beam waist for acceptable focal spot size}
When $w_0$ is equal to the radius of the entrance pupil $R$, the resulting theoretical $x$-FWHM values are larger by about 10\% for all three NA's than the ultimate FWHM values, which occur when $w_0 \gg R$. For NA=0.4, 0.75 and 0.95, $R$=4.0 mm, 3.0 mm and 1.71 mm, respectively. The difference between the theoretical $z$-FWHM's and  the ultimate $z$-FWHM's are 7.3\%, 4.8\% and 3.5\% for NA=0.4, 0.75 and 0.95, respectively, in this case.
When $w_0=R/2$, the difference increases to about 50\% for $x$-FWHM for all three NA's and to 88\%, 65\% and 54\% for NA=0.4, 0.75 and 0.95, respectively, for $z$-FWHM. In other words, the FWHM of the Gaussian beam at the focus does not increase much until the input beam waist is reduced below the half of the entrance pupil radius of the objective lens. Our experimental results support this observation.

\subsection{Effect of quasi-Gaussian input beam}

There exists a small discrepancy between theory and experiment in the regime of $w_0 \lesssim R/2$. It  is attributed to the use of an imperfect Gaussian beam as an input beam. In real experiments, the laser beam is not a perfect Gaussian beam. In order to make it close to a true Gaussian beam, spatial filtering of the laser beam is performed. For most of laser applications, a single-pass spatial filtering is more than enough. We have found, however, in our experiment with $w_0\ll R$ a single-pass spatial filtering was far from adequate since in this case the entire beam profile of the input beam determines the field distribution in the focal region.

Although a spatial filtering process can eliminate most of the asymmetric structures in the spatial distribution of an incident beam, the resulting mode distribution tends to contain small side lobes of Airy disk type. Unless these side lobes are thoroughly eliminated by a succession of extensive spatial filtering, the resulting field distribution in the focal region becomes significantly broadened and distorted from the expected distribution of the true Gaussian beam.
 
According to our numerical simulations, the error in FWHM measurement induced by the imperfect Gaussian beam may amount to 30\% or more and the error is particularly considerable in the regime of $w_0 \ll R$. In our experiment, we have used three successive stages of spatial filtering in order to minimize any deviation from the true Gaussian beam. When only a single stage of spatial filtering was used, we observed about 30\% increase in FWHM's in most cases.

\section{Summary and Conclusions} \label{sec6}

We studied the intensity distribution in the region of focus when a linearly-polarized well collimated Gaussian beam with a waist of $w_0$ was incident on a high-NA objective lens with an entrance pupil radius of $R$. We first theoretically examined the transverse spot size $\Delta x_{\rm FWHM}$ and the longitudinal spot size $\Delta z_{\rm FWHM}$ near the focal plane for an arbitrary input beam waist $w_0$. We used the vectorial diffraction theory of Richards and Wolf and calculated a time-averaged Poynting vector in the near focus. We then derived approximate expressions for FWHM's for two limiting cases, $w_0\ll R$ and $w_0\gg R$, and for the latter the approximate expression is in the form of Fraunhofer diffraction result although the result is obtained for the non-paraxial case.

In experiments, we varied the initial $w_0$ for a given NA's of 0.4, 0.75 and 0.95 and measured $\Delta x_{\rm FWHM}$ and $\Delta z_{\rm FWHM}$ values. They were obtained by scanning a pinhole of 0.5 $\mu$m diameter across the focused beam and by measuring the total transmitted light through the pinhole. The results obtained by convoluting the calculated Poynting vector with the pinhole were well matched with the measured intensity distributions. The smallest measured $x$- and $z$-FWHM values were 0.40 $\mu$m and 1.03 $\mu$m, respectively, for NA=0.95 with $\lambda$= 632.8 nm.

For high NA's with resulting focal beam waists comparable to or smaller than $\lambda$, observed $x$-FWHM's were smaller than those of the energy flux distribution convoluted with the pinhole. This discrepancy is attributed to the distortion of the electric field near the conducting surface of the pinhole. In addition, we observed a small discrepancy between theory and experiment for $w_0<R/2$, which is caused by slight deviation of the spatial distribution of the incident beam from that of the true Gaussian beam. Extensive multi-stage spatial filtering was used to minimize this deviation. 

Finally, we found both theoretically and experimentally that the FWHM of the Gaussian beam at the focus does not increase much until the input beam waist is reduced below the half of the entrance pupil radius of the objective lens. This result can be used as a practical design guideline for scanning microscopy employing a Gaussian laser beam as a probe. It has been also noted that the spatial distribution of the incident beam has to be as close to that of a true Gaussian beam as possible via extensive spatial filtering, in order to achieve the smallest focal beam spot, particularly when a beam with $w_0 \lesssim R/2$ is used as a scanning probe.

\section*{Acknowledgments}
This work was supported by Korea Research Foundation Grants (KRF-2002-070-C00044 and KRF-2005-070-C00058).



\begin{thebibliography}{99}

\bibitem{Marshall91} G.\ F.\ Marshall, ed., {\em Optical Scanning} (Marcel-Dekker, New York, 1991).

\bibitem{Belland82} P.\ Belland and J.\ P.\ Crenn, ``Changes in the characteristics of a Gaussian beam weakly diffracted by a circular aperture'', Appl. Opt. {\bf 21}, 522 (1982)

\bibitem{Tanaka85} K.\ Tanaka, N.\ Saga, and K.\ Hauchi, ``Focusing of a Gaussian beam through a finite aperture lens'', Appl.\ Opt.\ {\bf 24}, 1382 (1985)

\bibitem{Wolf59} E.\ Wolf., ``Electromagnetic Diffraction in Optical Systems. I. An Integral Representation of the Image Field'', Proc.\ R.\ Soc.\ London A {\bf 253}, 349 (1959).

\bibitem{Richards59} B.\ Richards and E.\ Wolf,  ``Electromagnetic Diffraction in Optical Systems. II.Structure of the Image Field in an Aplanatic System'', Proc.\ R.\ Soc.\ London A {\bf 253}, 358 (1959).

\bibitem{Mansuripur89} M.\ Mansuripur, ``Certain computational aspects of vector diffraction problems'', J.\ Opt.\ Soc.\ Am.\ A {\bf 6}, 786 (1989).

\bibitem{Born99} M.\ Born  and E.\ Wolf, {\em Principles of optics}, 7th (expanded) Ed. (Cambridge University Press, Cambridge, 1999).

\bibitem{Stamnes86} J.\ J.\ Stamnes, {\em Waves in Focal Regions} (Adam Hilger, Bristol and Boston, 1986).

\bibitem{Schneider81} M.\ B.\ Schneider and W.\ W.\ Webb, ``Measurement of submicron laser beam radii'', Appl.\ Opt.\ {\bf 20}, 1382 (1981)

\bibitem{Firester77} A.\ H.\ Firester, M.\ E.\ Heller, and P.\ Sheng, ``Knife-edge scanning measurements of subwavelength focused light beams'', Appl. Opt. {\bf 16}, 1971 (1977)

\bibitem{Quabis01} S.\ Quabis, R.\ Dorn, M.\ Eberler, O.\ Glockl and G.\ Leuchs, ``The focus of light -theoretical calculation and experimental tomographic reconstruction'', Appl.\ Phys.\ B  {\bf 72}, 109 (2001).

\bibitem{Dorn03} R.\ Dorn, S.\ Quabis, and G.\ Leuchs, ``The focus of light-linear polarization breaks the rotational symmetry of the focal spot'', J. Mod.Opt. {\bf 50},1917 (2003)

\bibitem{Rhodes89} S.\ K.\ Rhodes, A.\ Barty, A.\ Roberts and K.\ A.\ Nugent, ``Sub-wavelength characterization of optical focal structures'', Opt.\ Commun.\ {\bf 145}, 9 (1989).

\bibitem{Rhodes02} S.\ K.\ Rhodes, K.\ A.\ Nugent and A.\ Roberts, ``Precision measurement of the electromagnetic fields in the focal region of a high-numerical-aperture lens using a tapered fiber probe'', J.\ Opt.\ Soc.\ Am.\ A {\bf 19}, 1689 (2002).

\bibitem{mansuripur02} M.\ Mansuripur, {\em Classical Optics and its Applications} (Cambridge University Press, Cambridge, 2002).

\bibitem{E-field-resp} B.\ Sick and B.\ Hecht, and L.\ Novotny, ``Orientational Imaging of Single Molecules by Annular Illumination'', Phys.\ Rev.\ Lett.\ {\bf 85},4482 (2000).
\end{thebibliography}
\end{document}